\documentstyle[sprocl]{article}

\input{psfig}

\def\be{\begin{equation}}
\def\ee{\end{equation}}
\def\bea{\begin{eqnarray}}
\def\eea{\end{eqnarray}}

\begin{document}

\title{The gluon condensate from gauge invariant vortex vacuum texture } 

\author{Kurt Langfeld}

\address{Institut f\"ur Theor. Physik, Universit\"at T\"ubingen, 
Auf der Morgenstelle 14 \\ D-72076 T\"ubingen, Germany 
\\E-mail: langfeld@alpha8.tphys.physik.uni-tuebingen.de }


\maketitle\abstracts{ 
In $SU(2)$ lattice gauge theory, a new self-restricted cooling
procedure is developed to uncover the gauge invariant 
vortex vacuum texture. The emerging vortex vacuum 
structure amounts to the full string tension and gives rise 
to a mass dimension four condensate which is of pure vortex origin.
}

\section{Introduction} 

\noindent 
A revival of the vortex picture of quark confinement 
arose with the construction of the 
$p$-vortices which are defined after adopting the so-called center 
gauge~\cite{deb98} by projecting the gauge fixed link variables onto 
center elements~\cite{deb98}. In this case, it was observed that the 
vortices are 
sensible degrees of freedom in the continuum limit~\cite{la98}: 
the (area) density of the $p$-vortices as well as their binary interactions 
extrapolate to the continuum~\cite{la98}. The $p$-vortex picture of the 
Yang-Mills ground state also provides an appealing explanation of the 
deconfinement phase transition at finite temperatures~\cite{la99}. 
Recently, it was pointed out that the center gauge fixing which is prior 
to identify the physical vortex structure might be plagued by 
a so-called practical Gribov problem~\cite{kov99}. For this reason, as 
well as for rating the phenomenological importance of the vortices, 
a gauge invariant definition of the vortex vacuum texture is highly 
desired.  

In this talk, I propose a method for obtaining gauge invariant vortex 
structures and  investigate whether there is a genuine relation 
between these structures and the gluon condensate in SU(2) Yang-Mills 
theory. Extensive results and an adequate referencing  can be found in a 
subsequent publication~\cite{la00a}.

\section{Vortices, gluons and vortex induced condensates } 

In order to exhibit the degrees of freedom responsible for confinement
in SU(2) lattice Yang-Mills according to the $Z(2)$ vortex mechanism,
fractionizing $SU(2) \hat{=} Z_2 \times SO(3)$ is instrumental. 
The corresponding ''coordinates'' are center vortices and coset fields. 
Since the $SO(3)$ coset fields are isomorphic to algebra valued fields, 
these degrees of freedom are identified with the {\it gluonic} 
ones.

A new self--restricted cooling algorithm which reduces the $SO(3)$
action of the coset fields facilitates the gradual removal of the 
gluon fields from the lattice configurations while preserving the 
center degrees of freedom. For these purposes, I define a gluonic
action density per link by
\be
s^{gl}_{x,\mu} \; = \; \sum_{\bar{\nu } \not=  \pm \mu } \left\{
1 \; - \; \frac{1}{3} \, \hbox{tr} _A \,
O_{x,\mu \bar{\nu } } \right\} \; = \; \frac{1}{3}
\sum _{\bar{\nu } \not= \pm \mu } F^a_{\mu \bar{\nu }} [A]
~F^a_{\mu \bar{\nu } }[A]~a^4 \, + \, {\cal O}(a^6) \; ,
\label{eq:1}
\ee
where $O_{x,\mu\nu}$ is the plaquette calculated in terms of
the $SO(3)$ part of the $SU(2)$ link elements. 
The sum over $\bar{\nu}$ runs from $-4 \ldots 4$.
$F^a_{\mu\nu}[A]$ is the (continuum) field strength functional of the 
(continuum) gluon fields $A_\mu(x)$ and $a$ is the lattice spacing.
A local cooling step amounts for minimizing the action density 
(\ref{eq:1}) with respect to the coset fields at a given link. 
A self-restriction is imposed by rejecting the cooling of the adjoint 
link iff the gluonic action is smaller than some threshold value
$
s^{gl}_{x,\mu} \; < \; 8 \kappa ^4 \, a^4 \; .
$
Thereby $\kappa $ is a gauge invariant cooling scale of mass
dimension one. For $\kappa=0$, the cooling procedure completely removes
the gluon fields from the SU(2) lattice configurations leaving only
center degrees of freedom. For $\kappa $ smaller than the string tension, 
a clear signal of the vortex vacuum texture is noticed by a large 
SU(2) action density which is accumulated at the two dimensional vortex
world sheets of the 4-dimensional space time. Since the new cooling 
procedure is gauge covariant and since the vortex structure manifests 
itself in the gauge invariant $SU(2)$ action density, the 
vortex texture is gauge invariant~\cite{la00a}. 

In order to study the relevance of this vortex texture, 
I calculated the force between a static quark anti-quark pair 
from $SO(3)$ cooled configurations. 
\begin{figure}[t]
\rule{5cm}{0.2mm}\hfill\rule{5cm}{0.2mm}
\psfig{figure=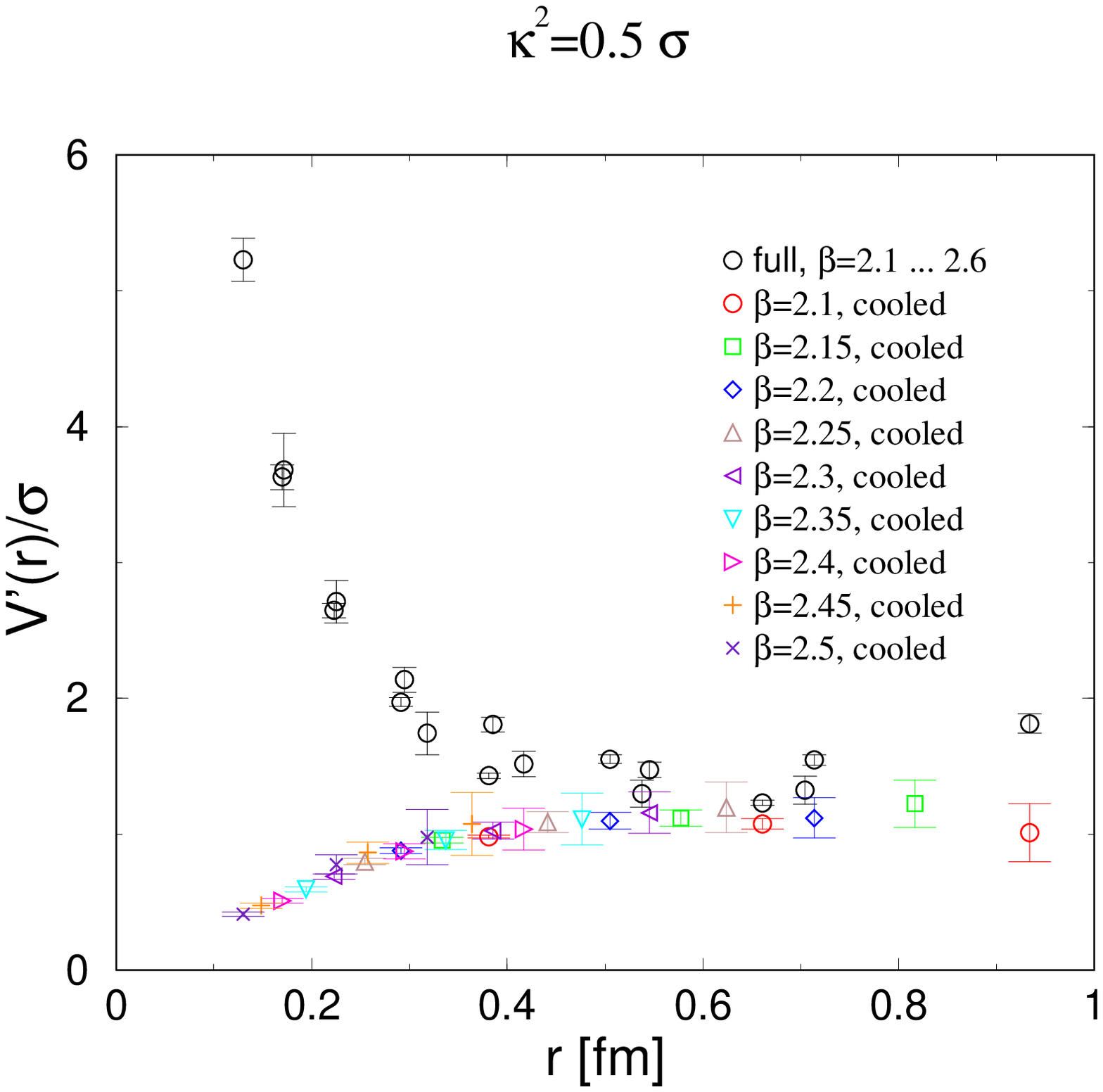,width=2.1in,height=1.5in}
\psfig{figure=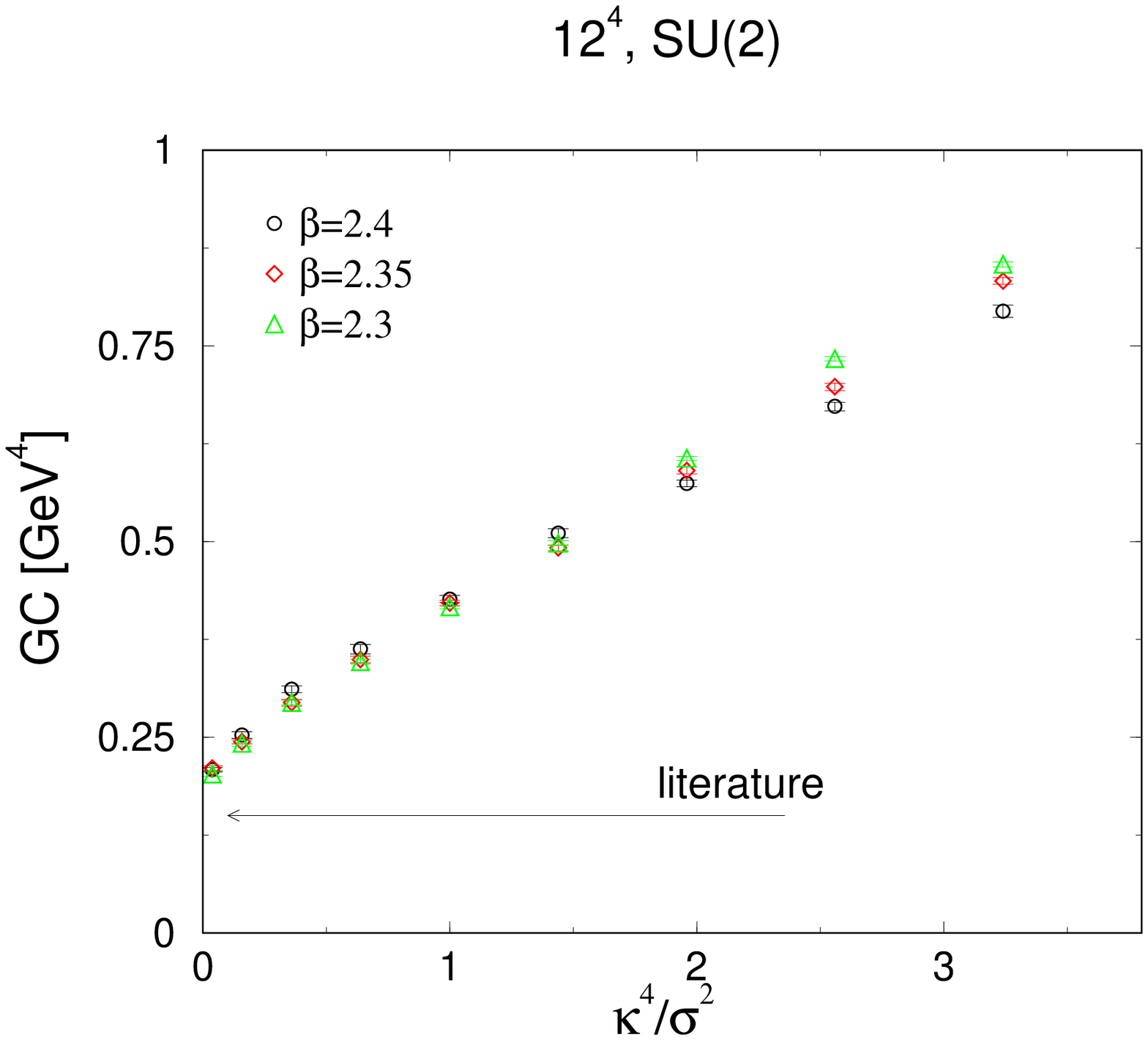,width=2.1in,height=1.5in}
\rule{5cm}{0.2mm}\hfill\rule{5cm}{0.2mm}
\caption{The static quark $Q\overline{Q}$ force as function
  of the distance $r$ between quark and anti--quark for
  full non--Abelian and $SO(3)$ cooled configurations. 
\label{fig:1}}
\end{figure}
I find that the cooling
procedure has strongly affected the force at short distances.
This is expected since the behavior at small $r$ is dominated by
the exchange of gluons, which are already partially eliminated by cooling.
Most important is, however, that the string tension is unchanged by the 
cooling procedure.

Hadronic correlators at high momentum transfer are accessible by experiments. 
On the other, these correlators are sufficiently under the control 
of the operator product expansion (OPE), where non--perturbative 
properties of the Yang--Mills (or QCD) vacuum are parameterized by 
so--called condensates~\cite{shif79}. The mass dimension four 
operator, $GC$, is provided by the trace of the energy momentum tensor, and, 
hence proportional to the $SU(2)$ action density in the present case. 
Here, I will use the new cooling 
method to uncover the contribution of the vortices
to the mass dimension four condensate of $SU(2)$ Yang--Mills theory.
It turns out, that the vortex contribution to 
$GC$ shows appropriate scaling towards the 
continuum limit~\cite{la00a}. The mass dimension four condensate 
$GC$ in physical units is also shown in figure~\ref{fig:1} as function 
of $\kappa ^4$. At large values of $\kappa $, the contribution of the 
coset, say gluon, fields to the condensate is large and $GC$ linearly 
rises with $\kappa ^4$. However, the most striking feature is that 
$GC$ approaches a finite value in the limit $\kappa \rightarrow 0$, thus 
showing a non-trivial dimension four condensate which is of pure 
vortex origin. The limiting value of $0.19 \, \hbox{GeV}^4$, is 
in rough agreement with recent values for the gluon condensate 
quoted in the literature~\cite{ilg00}. 

\section*{Acknowledgments}
It is a pleasure to thank my collaborators E.-M.~Ilgenfritz and 
H.~Reinhardt.

\end{document}